\documentclass[12pt,a4paper]{article}

\usepackage{graphicx}
\usepackage[justification=centering,font={small}]{caption} 
\usepackage{subcaption} 
\graphicspath{{./figures/}}

\usepackage{natbib}
\usepackage{amsmath}
\usepackage{anysize}
\usepackage{multirow}
\usepackage[usenames,dvipsnames]{xcolor}

\usepackage{titlesec}
\titleformat*{\section}{\large\bfseries}

\marginsize{2cm}{2cm}{2cm}{2cm} 
\newcommand{\ud}{\,\mathrm{d}}

\title{\Large Energetics of mixing for the filling box and the emptying-filling box} 

\author{\normalsize Megan S. Davies Wykes, Charlie Hogg, Jamie Partridge and Graham O. Hughes}

\date{\normalsize \today}

\begin{document}

\maketitle

\begin{abstract}
	The mixing efficiency of a plume in a filling box and an emptying-filling box is calculated for both transient and steady states. The mixing efficiency of a plume in a filling box in an asymptotic steady state is 1/2, independent of the details of this state or how the plume is modelled. The mixing efficiency of a plume in an emptying filling box in steady state is $1 - \xi$, where $\xi = h/H$, the depth of the ambient layer $h$ non-dimensionalised by the height of the box $H$. A deeper mixed layer therefore corresponds to a higher mixing efficiency.
\end{abstract}

\section{Introduction}

The characterisation and measurement of turbulent stratified mixing remains a central problem for developing models of oceans, lakes and the atmosphere \cite{Lozovatsky2013}. Mixing in stratified flows can be quantified in a variety of ways. One common means to describe how the density field is modified in a free turbulent flow, such as a plume, jet, or gravity current, is through use of an entrainment coefficient that models the incorporation of ambient fluid into the turbulent region. For a plume, the entrainment coefficient $\alpha = u_e/U$, where $u_e$ is a horizontal velocity scale for ambient fluid drawn into the plume and $U$ is a vertical velocity scale for fluid in the plume \cite{Morton1956}. This description of the turbulent entrainment is suggested by dimensional analysis and has been used successfully on plumes with a wide range of length scales \cite{Linden1999}. The entrainment coefficient for plumes has recently been linked to the production of turbulent kinetic energy \cite{VanReeuwijk2015} and buoyancy variance \cite{Craske2017}, tying the entrainment coefficient to both viscous dissipation and irreversible mixing.

Another measure of mixing in a stratified flow is based on the fact that mixing of a stratification modifies the gravitational potential energy budget. Increases in potential energy in a stratified flow can be reversible (e.g.\ in an internal wave) or irreversible (e.g.\ when two parcels of fluid mix, changing the density of both), but only the irreversible increases correspond to mixing. The most common framework for differentiating between irreversible and reversible changes in potential energy splits the gravitational potential energy into available potential energy and background potential energy \cite{Lorenz1954,Winters1995}. Stratified mixing can then be characterised by the mixing efficiency, which compares the energy used in irreversible diabatic mixing to the energy that was available for mixing \cite{Peltier2003}. In this paper, we use an energetics framework to examine mixing in the filling box and the emptying-filling box.

The background potential energy, $E_b$, is the gravitational potential energy of the system if every parcel of fluid were allowed to rise or fall without changing its density until the system reaches a state of minimum gravitational potential energy. This minimum potential energy is equivalent to the potential energy of the fluid volume if the rearranged density profile increases monotonically in the direction of the gravitational vector \cite{Lorenz1954}. In a closed system, the gravitational potential energy of this rearranged profile can only increase as a result of mixing, i.e.\ changing the density of fluid parcels, which raises the centre of mass of this reference state. As mixing is irreversible, for an open system in steady state, net buoyancy fluxes across the boundaries must result in reduction of the background potential energy at the same rate that mixing increases the background potential energy within the system.

The available potential energy, $E_a$, of a given state is the energy that would be released by the above rearrangement of fluid parcels and is energy that is available to do work in the system. It is non-zero when the profile is not in a state of gravitational equilibrium, such as if kinetic energy in the flow moves a parcel of fluid away from its height of neutral buoyancy. Available potential energy can also be added (removed) directly to (from) the flow by introducing buoyancy forcing or advection of fluid across the boundaries of the system.

The energy used in mixing can be measured using the background potential energy, ${E}_b$. The energy available for mixing is the sum of kinetic energy, ${E}_k$, and available potential energy, ${E}_a$, present in the system. As we will consider an unsteady flow in this paper, we make use of the instantaneous mixing efficiency, which can be expressed as
\begin{equation}
	\eta = \frac{\dot{E}_b}{\left|\dot{E}_a + \dot{E}_k\right|} \label{eq:mixing_efficiency}
\end{equation}
for a closed system, where $\dot{E}_b$ is the rate of change of background potential energy (positive when irreversible mixing takes place) and $\dot{E}_a$ + $\dot{E}_k$ is rate of supply of available energy. For a closed system, $\dot{E}_b$, $\dot{E}_a$ and $\dot{E}_k$ represent rates of conversion within the system. For cases where mass or energy transfers across the boundaries are permitted, these represent sources or sinks of energy that must also be taken into account.

In situations where the mixing is buoyancy-driven (i.e.\ the source of energy in the system is initially entirely ${E}_a$), high mixing efficiencies have been observed, with values of the cumulative mixing efficiency greater than $75\%$ measured in experiments of Rayleigh--Taylor instability \cite{DaviesWykes2014}, and tending towards $100\%$ in experiments of horizontal convection \cite{Gayen2013a}. In some cases of buoyancy-driven stratified mixing, the mixing efficiency depends on the density profile in regions remote from where the mixing takes place \cite{DaviesWykes2015}. Nevertheless, the mixing efficiency is often discussed in the literature as if it were a constant value or some property of the turbulence itself, therefore it seems useful to examine and understand cases where this is not true, to extend our intuition about such cases. The filling box and the emptying-filling box are both simple, well-defined  systems in which the steady state dynamics are well understood, making them useful test-cases.

A simplified model of a plume within a closed container filled is known as a `filling box'. Without loss of generality, we will assume that the plume originates from a point source of pure buoyancy and falls through a box that is of height $H$. As a result of entrainment into the plume, a stable stratification will develop in the box \cite{Baines1969}. This stable stratification can be predicted using plume theory, which assumes that entrainment of fluid from the ambient into the plume at some height is proportional to the mean vertical velocity at that height \cite{Morton1956}. Time-dependent density profiles for a plume in a box have also been derived, along with approximate analytic expressions for the density profile \cite{Worster1983}.

\begin{figure}[ht!]
\centering
\includegraphics[width=0.45\textwidth]{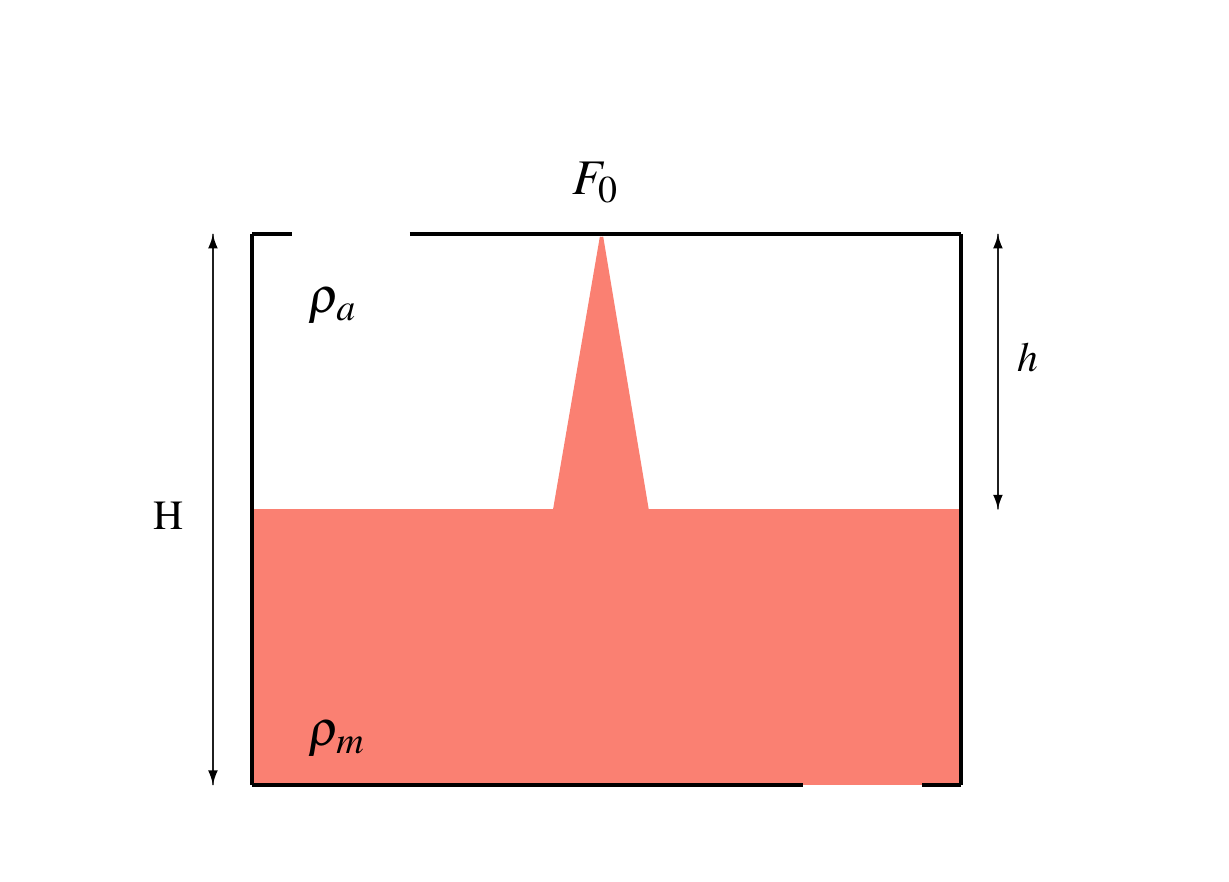}
\caption{Steady state of the emptying-filling box with a point source of pure buoyancy (whose specific buoyancy flux is $F_0$). $H$ is the vertical separation between the openings, $h$ is the distance from the upper opening to the interface. The buoyant mixed layer and ambient densities are labelled $\rho_m$ and $\rho_a$ respectively.}\label{fig:emptying_filling_diagram}
\end{figure}

The emptying-filling box (Fig.\ \ref{fig:emptying_filling_diagram}) is a conceptual extension of the filling box, which introduces openings through the top and bottom \cite{Linden1990}. There is a resulting pressure difference between the openings that drives a ventilation flow. Ambient fluid enters the box through the opening at the top and mixed fluid exits through the opening at the bottom. The emptying-filling box can reach a steady state when the removal of dense fluid by the ventilation flow is balanced by addition of dense fluid to the mixed layer by the plume. For a single source of buoyancy, the steady state consists of a layer of ambient fluid of thickness $h$ and density $\rho_a$ that sits above a dense mixed layer, which has constant density $\rho_m > \rho_a$. The depth of the mixed layer within the box is independent of the magnitude of the source buoyancy flux and controlled only by the entrainment coefficient $\alpha$ of the plume and a ratio of the height of the box to a quantity $A^*$, which is a function of the areas of the two openings \cite{Linden1990}.

The emptying-filling box is a common model for buoyancy-driven natural ventilation and has been used to study steady states \cite{Linden1990} and transients \cite{Kaye2004,Bower2008,Sandbach2011}. Natural ventilation of buildings makes use of wind or temperature differences to drive ventilation flow, rather than mechanical forcing \cite{Linden1999}. Displacement ventilation uses existing buoyancy sources to draw cool, fresh air into a building through an opening near the floor, while removing warm air through an opening near the ceiling. Note that in the example shown in Fig.\ \ref{fig:emptying_filling_diagram}, the plume is dense and falls into a less dense ambient, whereas in natural ventilation warm plumes would rise through a cool ambient.

In this paper, we examine how the mixing efficiency and entrainment coefficient associated with a plume in a filling box and an emptying-filling box relate to the turbulent mixing that takes place. We consider both the transient and steady states in the flow. 

\section{The Filling Box}

\subsection{Mixing efficiency of the asymptotic steady state}

The filling box reaches an asymptotic steady state \citep{Baines1969}, in which the density gradient in the interior is a function of position $z$ only, and the rate of change of density is both spatially uniform and constant, i.e.\ $\partial \rho(z,t)/\partial t$ is constant. Thus irreversible mixing in the filling box arranges itself so that the rate of change of density is uniform in space.

If the plume has specific buoyancy flux $F_0$ (m$^4$s$^{-3}$) and the room has constant cross-sectional area, then the rate of change of density at all heights in the basin is
\begin{equation}
\frac{\partial \rho}{\partial t} = \frac{\hat{\rho}F_0}{gV}
\end{equation}
where $\hat{\rho}$ is a reference density, $g$ is gravitational acceleration and $V$ is the volume of the filling box. If we assume the volume of the plume is small compared with $V$, and that fluid in the filling box is everywhere close to its equilibrium level, then the rate at which the potential energy of fluid in the box (i.e.\ the background potential energy) changes owing to irreversible mixing is given by
\begin{equation}
\dot{E}_b = g\int_{V} \frac{\partial \rho}{\partial t} z \ud V = \frac{\hat{\rho} F_0 H}{2}.
\end{equation}

The buoyancy forcing that maintains the plume is the only source of energy for the filling box. The rate of addition of available potential energy by the plume source is equal to the change in potential energy if the buoyancy released by the plume traversed the depth of the box without mixing, a process equivalent to sorting the unstable density profile. The rate at which available potential energy is supplied to the system by the buoyancy source is  
\begin{equation}
	\dot{E_a} = \hat{\rho} F_0 H, \label{eq:APE}
\end{equation}
where $F_0$ is the buoyancy flux at the plume source, and $\hat{\rho}$ is a reference density. This buoyancy forcing results in parcels of fluid close to the plume source that are positively or negatively buoyant, i.e.\ they have available potential energy. These parcels of fluid pass through the box depth, converting available potential energy to kinetic energy in the process. Turbulence and density gradients arise on small scales, and energy is consumed by irreversible mixing and viscous dissipation.

The turbulent mixing efficiency $\eta$ for the asymptotic steady state of a filling box can thus be estimated as the ratio of the rate of irreversible mixing, $\dot{E_b}$, and the rate of release of available potential energy,
\begin{equation}
\eta = \frac{1}{2}. \label{eq:eta_filling_box} 
\end{equation}
In using this result to characterise turbulent mixing in the filling box, it is assumed that the kinetic energy dissipated from the mean overturning flow is negligible (i.e.\ $\dot{E}_k \ll \dot{E}_a$). The contribution to irreversible mixing associated with molecular diffusion down the mean background gradient through the box depth is also assumed to be unimportant.

Equation \ref{eq:eta_filling_box} is an interesting result in that the mixing efficiency of the asymptotically steady state is not a function of the entrainment coefficient or any other parameter. It is also independent of the model used to describe the plume and only requires that an asymptotic steady state is reached, without depending on any of the details of that state. A mixing efficiency of $1/2$ has been found in other flows \cite{Dalziel2008b} where mixing is driven by available potential energy -- consistent with the maximum mixing efficiency for any 1D monotonic unstable stratification, which has been shown to be $1/2$ \citep[see][Appendix A]{DaviesWykes2015}.

\subsection{Mixing efficiency of the transient state}
\label{sec:filling_box_model}

We can also examine the time dependent evolution of an axisymmetric plume in a box \cite{Baines1969}. An axisymmetric plume is maintained below this point source and, at any height $z$, the vertical velocity and buoyancy are assumed to have mean Gaussian profiles, where $w$, $F$ and $b$ are the maximum vertical velocity, maximum buoyancy and Gaussian half-width, respectively. As in our last example, we consider a dense plume in a less dense ambient.

\begin{figure}[ht!]
	\centering
	
	\begin{subfigure}[t]{0.45\textwidth}
		\centering
		\includegraphics[width=\textwidth]{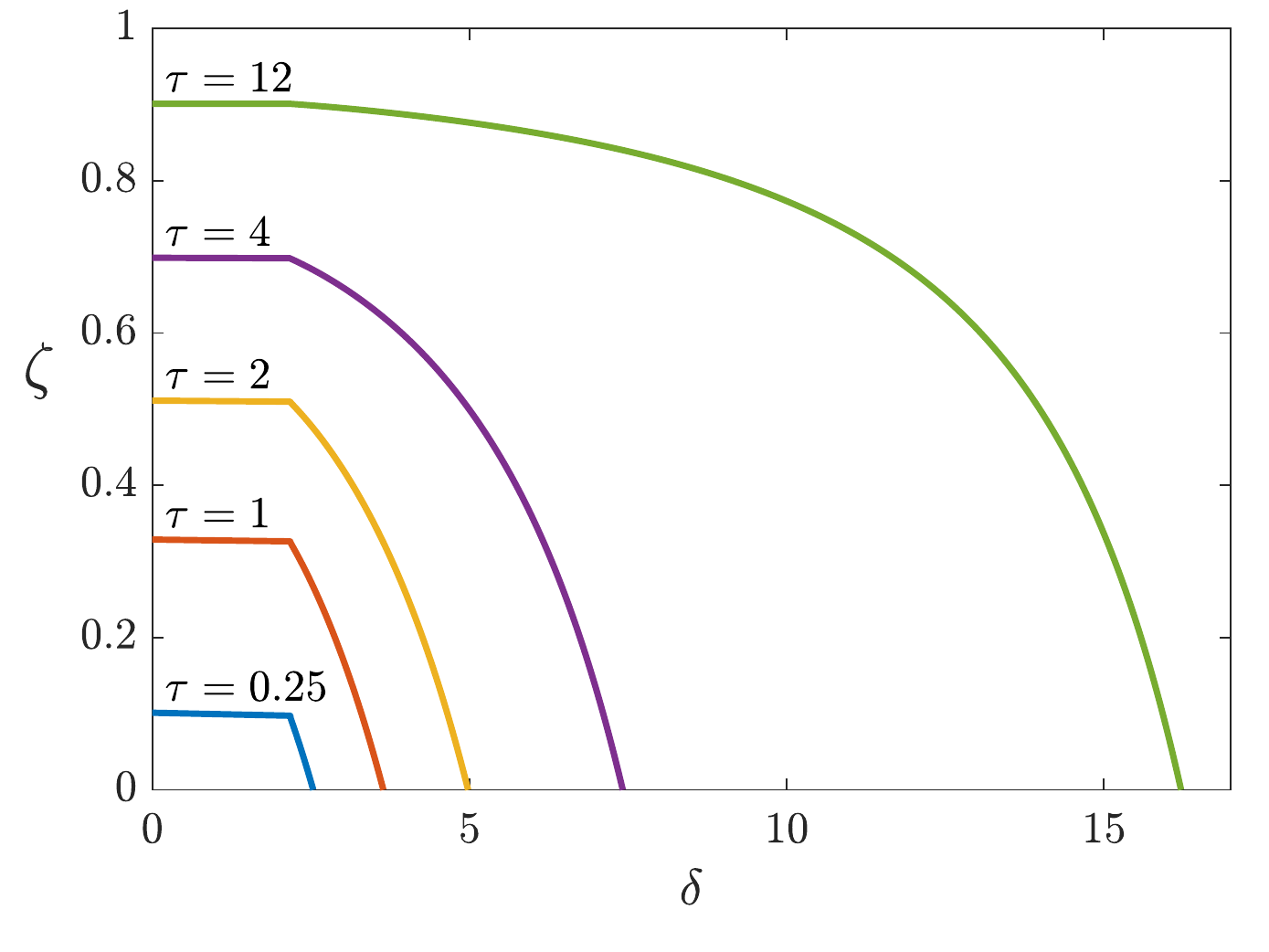}
		\caption{} \label{fig:fillingbox_density}
	\end{subfigure}%
	~ 
	\begin{subfigure}[t]{0.45\textwidth}
		\centering
		\includegraphics[width=\textwidth]{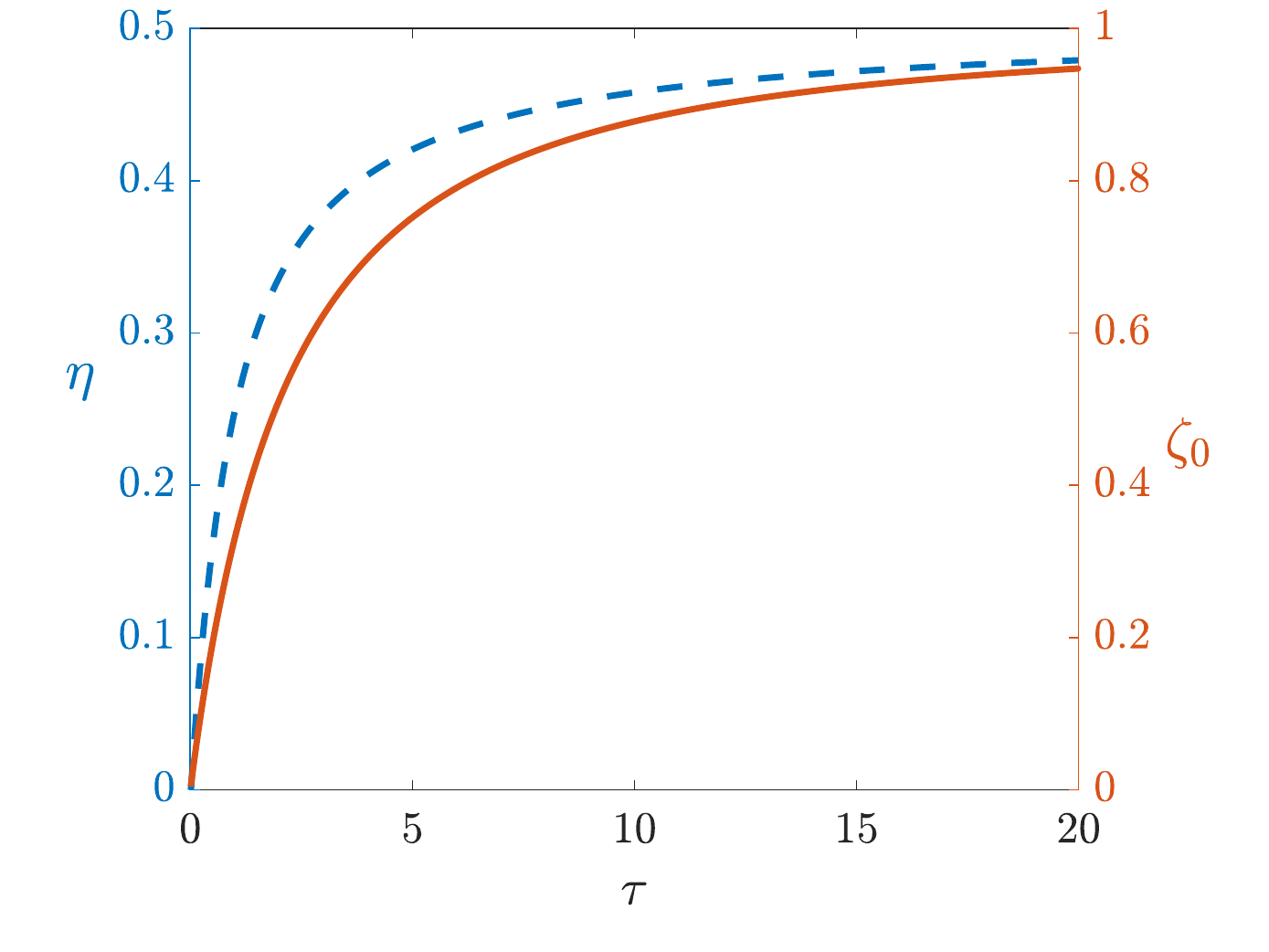}
		\caption{} \label{fig:fillingbox_efficiency}
	\end{subfigure}
	
	\caption{Filling box: (a) Dimensionless density profile at $\tau = 0.25, 1, 2, 4, 12$, (b) mixing efficiency $\eta$ (blue, dashed) and the height of the mixed layer $\zeta_0$ (orange, solid) against time non-dimensionalised by a filling box time scale, as defined in Eq.\ \ref{eq:nondim_variables}.}
\end{figure}

The equations that govern the time evolution of the density profile as the box fills are given by Worster and Huppert, who also calculated an approximate analytical solution \cite{Worster1983}. We follow their analysis to compute the density profile in the tank as a function of height and time $\rho(z,t)$. The dimensionless quantities for height, density, time, buoyancy flux, volume flux, and momentum are defined as 
\begin{equation}
\begin{aligned}
	\zeta = z H^{-1}, \qquad &
	\delta = 4 \pi^{\frac{2}{3}} \alpha^{\frac{4}{3}} H^{\frac{5}{3}} F_0^{-\frac{2}{3}} g \frac{\rho - \hat{\rho}}{\hat{\rho}},  \\
	\tau = 4 \pi^{\frac{2}{3}} \alpha^{\frac{4}{3}} H^{\frac{2}{3}} A^{-1} F_0^{\frac{1}{3}} t, \qquad &
	f = \frac{1}{2} \pi F_0^{-1} b^2 w F, \\
	q = \frac{1}{4} \pi^{\frac{1}{3}} \alpha^{-\frac{4}{3}} H^{-\frac{5}{3}} F_0^{-\frac{1}{3}} b^2 w, \qquad &
	m = \frac{1}{2} \pi^{\frac{1}{3}} \alpha^{-\frac{1}{3}} H^{-\frac{2}{3}} F_0^{-\frac{1}{3}} b w.
\end{aligned} \label{eq:nondim_variables}
\end{equation}
The non-dimensionalised governing equations are
\begin{equation}
	\frac{\ud q}{\ud \zeta} = m, \qquad
	\frac{\ud {m^2}}{\ud \zeta} = \frac{q f}{m^2}, \qquad
	\frac{\ud f}{\ud \zeta} = q \frac{\partial \delta}{\partial \zeta}, \qquad
	\frac{\partial \delta}{\partial \tau} = q \frac{\partial \delta}{\partial \zeta}. \label{eq:nondim_equations}
\end{equation}
These equations conserve the fluxes of volume, momentum, and buoyancy in the plume, respectively, and describe the evolution of the density profile in the ambient. We solve Eqns.\ \ref{eq:nondim_equations} numerically using a layered Germeles model and a second order Runge-Kutta scheme \cite{Germeles1975}. At each time-step, a layer is added to the bottom of the density profile, with the volume and density in the layer computed from the turbulent plume equations \citep[for more details see][]{Germeles1975,Sandbach2011}. The change in volume of other layers due to exchange with the plume is calculated from mass conservation and assuming that the volume taken up by the plume is much smaller than the volume of the box. The evolution of the dimensionless density profile is plotted in Fig.\ \ref{fig:fillingbox_density}, with the position of the upper edge of the mixing region $\zeta_0$, -- also known as the first front -- is plotted in Fig.\ \ref{fig:fillingbox_efficiency}.

To calculate the instantaneous mixing efficiency requires the rate of supply of available potential energy in the system and the rate of increase of background potential energy. In the non-dimensionalised problem all rates of energy transfer are in effect normalised by the rate of supply of available potential energy. Therefore the mixing efficiency is equal to the rate of increase of the normalised background potential energy, which is equal to the rate of increase of the normalised potential energy. We will assume the box volume is large compared to the volume taken up by the plume and therefore neglect the contribution of fluid in the plume to the potential energy budget.

The evolution of the mixing efficiency as the box fills is shown in Fig.\ \ref{fig:fillingbox_efficiency}. At early times the mixing efficiency is low as parcels of fluid that are mixed in the plume are dense and always fall to the stratified layer at the bottom of the box, where they transfer all their available potential energy into kinetic energy that is  dissipated viscously. The mixing efficiency increases monotonically with the height of the mixed layer, tending towards a maximum value of $1/2$. At late times, the increase in density of the box induced by the presence of the buoyancy source is equally distributed across the full depth. Another way of thinking about this is that a parcel of dense fluid introduced at the top of the box has an initial centre of mass $z = H$. This density change is equally distributed across the entire box, therefore the final centre of mass is $z = H/2$, or equivalently, half the initially available potential energy has been transformed by mixing into background potential energy.

The entrainment coefficient appears in the non-dimensionalisation of time so, although the mixing efficiency of the final state is independent of the entrainment coefficient, the time taken to reach the final state does depend on the entrainment coefficient. This suggests a view of the entrainment coefficient in this example as associated with a rate of mixing, rather than directly with the energetic consequences of mixing.

\section{The Emptying-Filling Box}

\subsection{Mixing efficiency of the steady state}

A plume model based on the use of an entrainment constant can be used to model the flow in an emptying-filling box. It has been shown, both theoretically and in laboratory experiments, that the emptying-filling box establishes a steady state \cite{Linden1990}. In the steady state, the depth of the upper layer satisfies
\begin{equation}
\left(\frac{\xi^{5}}{1-\xi}\right)^{1/2} = \frac{A^*}{C^{3/2}H^2}, \label{eq:ef_steady_state_layer_thickness}
\end{equation}
where $\xi = h/H$ is the dimensionless thickness of the ambient layer, $C = \pi \left(\frac{5}{2\pi\alpha}\right)^\frac{1}{3} \left(\frac{6 \alpha}{5}\right)^\frac{5}{3}$ for an axisymmetric plume with Gaussian profiles, $A^* = a_1a_2/\sqrt{\frac{1}{2}(a_1^2/c + a_2^2)}$, $a_1$ and $a_2$ are the areas of the upper and lower openings, and $c \approx 0.6$ is a constant associated with the loss coefficients of the two openings \cite{Kaye2004}. As the dimensionless opening area $A^*$ is reduced towards zero, the dimensionless depth of the ambient layer $\xi$ decreases. When $A^*$ is increased, the dimensionless depth of the ambient layer increases towards one.

The steady state reached in the emptying filling box has constant background potential energy, in contrast to the filling box. The rate of removal of background potential energy from the system by the flow through the box is exactly offset by irreversible mixing. This rate is therefore given by the rate of working by buoyancy forces if the density interface (Fig.\ \ref{fig:emptying_filling_diagram}) were to be advected by the through flow.

We calculate the rate of removal of potential energy from the system by considering the instantaneous rate of increase in potential energy if the through flow were momentarily halted (while holding the other parameters constant). In calculating this we neglect the volume of gravitationally unstable fluid in the plume and consider only the globally stable two-layer stratification in the box. The potential energy of the box at time $t$ is
\begin{equation}
	E_b(t) = \frac{g(\rho_m - \rho_a)(H-h)^2}{2} + \frac{g \rho_a H^2}{2},
\end{equation}
i.e.\ the mass multiplied by $g$ multiplied by the height of the centre of mass. If the ventilation flow were switched off, after a short time $\Delta t$, the background potential energy would be
\begin{equation}
	E_b(t + \Delta t) = \frac{g(\rho_m - \rho_a)(H-h + \Delta t \, Q)^2}{2} + \frac{g \rho_a H^2}{2},
\end{equation}
where $Q$ is the flow rate of fluid through the lower opening and $h$ is the depth of the ambient density layer.
The rate of removal of background potential energy from the system by the through flow is given by 
\begin{align}
E_b(t + \Delta t) - E_b(t) 
& = \frac{g \Delta t \, Q}{2}(\rho_m - \rho_a)(2(H - h) + \Delta t \, Q).
\end{align}
Dividing by $\Delta t$ and taking the limit $\Delta t \rightarrow 0$, gives the instantaneous increase in potential energy as
\begin{equation}
\dot{E_b} = g(\rho_m - \rho_a)(H - h)Q.
\end{equation}
As buoyancy is conserved, $g (\rho_m - \rho_a) Q = \hat{\rho} F_0$, therefore
\begin{equation}
\dot{E_b} = (H - h)\hat{\rho} F_0. \label{eq:BPE}
\end{equation}
This is the energy input required to maintain the height of the mixed layer against the action of the through flow.

The mixing efficiency is the ratio between the rate of increase in background potential energy and the rate at which available potential energy is supplied to the system (Eq.\ \ref{eq:mixing_efficiency}). Substituting the values for the background potential energy (Eq.\ \ref{eq:BPE}) and available potential energy (Eq.\ \ref{eq:APE}), we find
\begin{equation}
\eta = 1 - \xi, \label{eq:emptying_filling_mixing_efficiency}
\end{equation}
where $\xi = h/H$ is the non-dimensional depth of the ambient layer. If there is a thick mixed layer (i.e.\ a thin ambient layer, $\xi = h/H \rightarrow 0$), the mixing efficiency increases towards $1$, while if the mixed layer thickness decreases (i.e.\ the ambient layer depth increases), the mixing efficiency decreases towards zero.  

\begin{figure}[ht!]
\centering
\includegraphics[width=0.45\textwidth]{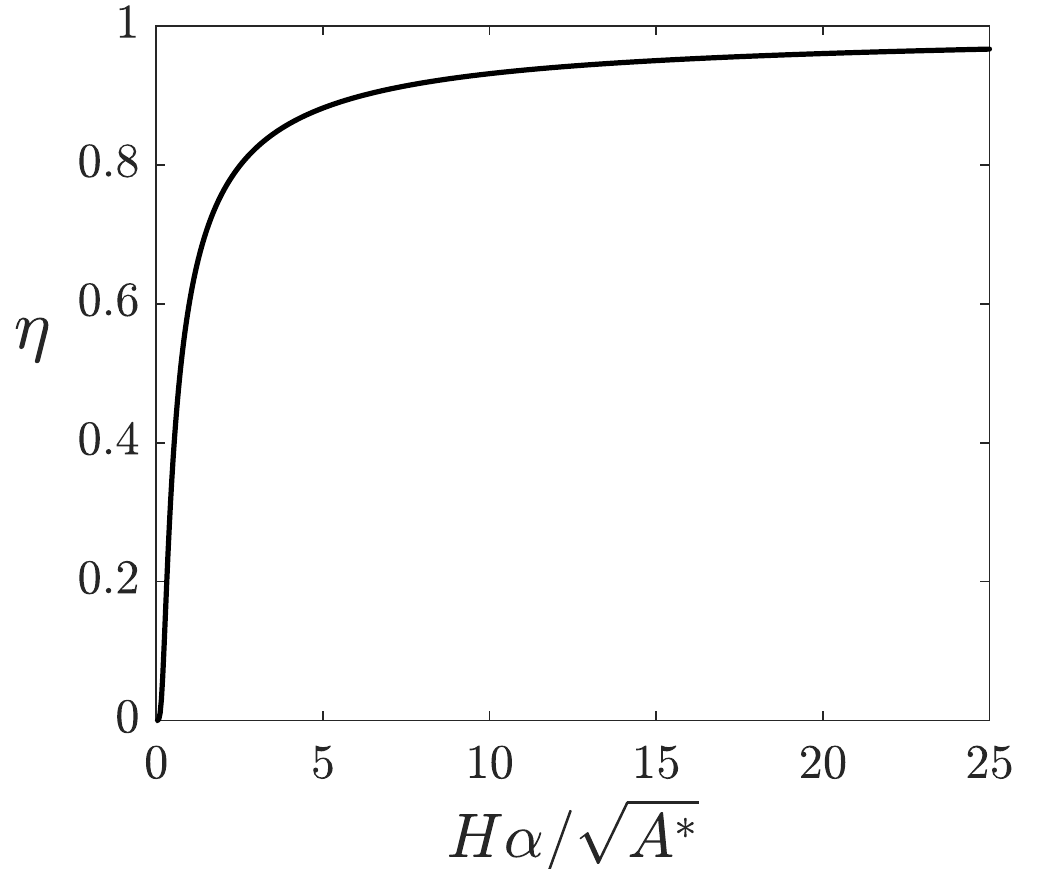}
\caption{Emptying-filling box in steady state: Mixing efficiency $\eta$, as a function of the opening area $A^*$, box height $H$, and entrainment coefficient $\alpha$.}\label{fig:emptyingfillingbox_mixing}
\end{figure}

By substituting for the layer depth in Eq.\ \ref{eq:emptying_filling_mixing_efficiency} and Eq.\ \ref{eq:ef_steady_state_layer_thickness}, the mixing efficiency may be expressed implicitly as a function of the opening area $A^*$, box height $H$ and the entrainment coefficient $\alpha$,
\begin{equation}
\frac{\eta}{(1 - \eta)^5} = \frac{1}{3 \pi^2} \left(\frac{5}{6}\right)^4 \frac{(H \alpha)^4}{A^{*2}} \label{eq:14}
\end{equation}
The behaviour of $\eta$ as a function of $\frac{H \alpha}{\sqrt{A^*}}$is plotted in Fig.\ \ref{fig:emptyingfillingbox_mixing}. When $A^*$ is reduced, the thickness of the mixed layer increases and the mixing efficiency increases towards $1$. Conversely, when $A^*$ is increased, the thickness of the mixed layer decreases and the mixing efficiency approaches zero.

We can understand this by considering the balance between available potential energy and kinetic energy in the flow. As the plume falls, fluid loses available potential energy to kinetic energy, until the plume reaches the mixed layer. When the plume enters the mixed layer little further mixing can occur as the mean density in the plume is equal to the density of the mixed layer. If the mixed layer is relatively thin the plume falls almost the entire height of the box before being arrested, losing almost all available potential energy to kinetic energy, which is then dissipated, resulting in a low mixing efficiency.

The entrainment coefficient is often considered as a constant for a plume, but in other flows such as inclined gravity currents over rough walls the entrainment coefficient is different from the value measured for plumes \cite{Fernandez2006}. As a thought experiment we can examine the effect on the mixing efficiency of varying the entrainment coefficient while holding $A^*/H^2$ constant. Decreasing the entrainment coefficient reduces the mixed layer depth and decreases the mixing efficiency. However, the value of the mixing efficiency is not determined by the entrainment coefficient, as the full range of values from 0 to 1 is theoretically possible by varying only $A^*$. 

\subsection{Mixing efficiency of the transient state}

We can create a simple model for the transient density profile in an emptying-filling box by adding an equation for the ventilation rate to our model from \S \ref{sec:filling_box_model} \cite{Bolster2008}. The ventilation rate is driven by the pressure difference between the top and bottom openings, which is determined by the density profile in the tank. In our model, the ventilation rate is given by 
\begin{equation}
Q_{v} =  A^* \left(\sum_{i=1}^n g'_i (z_{i} - z_{i-1}) \right)^{\frac{1}{2}},
\label{eq:ventilation_rate}
\end{equation}
where $z_i$ is the height of the $i$th layer and $g'_i$ is the reduced gravity \cite{Sandbach2011}. When we non-dimensionalise (as in Eq.\ \ref{eq:nondim_variables}), Eq.\ \ref{eq:ventilation_rate} becomes
\begin{align}
q_{v} & =  \frac{1}{8 \pi}\, \frac{A^*}{\alpha^2 H^2} \left(\sum_{i=1}^n \delta_i (\zeta_{i} - \zeta_{i-1}) \right)^{\frac{1}{2}} 
\label{eq:nond_ventilation_rate}
\end{align}
where $q_{v}$ is the non-dimensional ventilation rate. A similar model for the emptying-filling box was used by Sandbach and Lane-Serff, who used a modified version of the equation that links plume volume flow rate and layer heights in the ambient \cite{Sandbach2011}.

In our model, we calculate the effect of the plume on the stratification in the box as before (\S \ref{sec:filling_box_model}). The ventilation flow is included by shifting the stratification down by $\zeta_v = q_v \Delta \tau$ and then removing a layer of thickness $\zeta_v$ from the bottom. The potential energy of the box $E_{box}$ and the change in potential energy before and after the ventilation occurs are calculated at each time step.

\begin{figure}[ht!]
	\centering
	
	\begin{subfigure}[t]{0.45\textwidth}
		\centering
		\includegraphics[width=\textwidth]{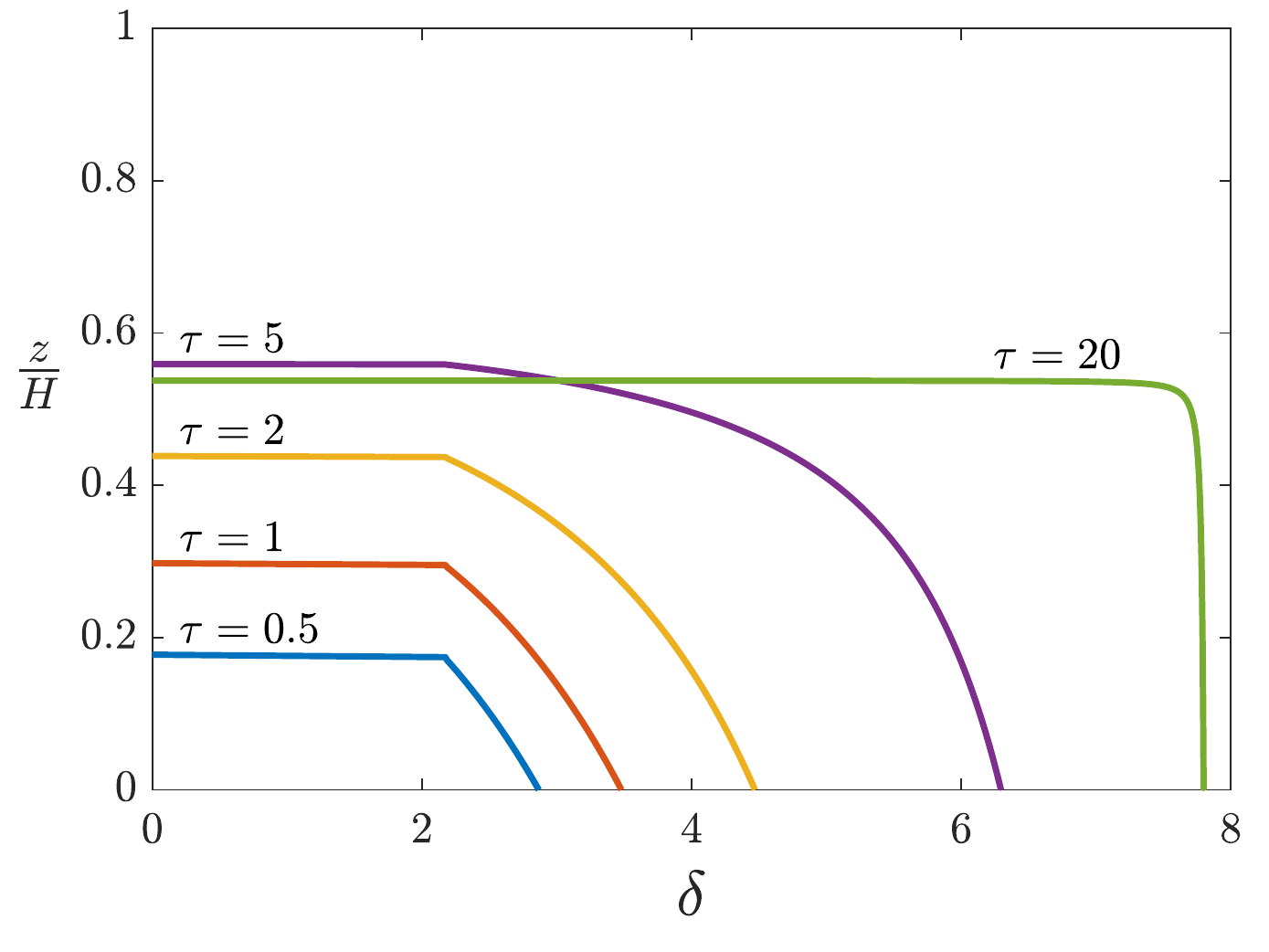}
		\caption{} \label{fig:emptyingfillingbox_density}
	\end{subfigure}%
	~ 
	\begin{subfigure}[t]{0.45\textwidth}
		\centering
		\includegraphics[width=\textwidth]{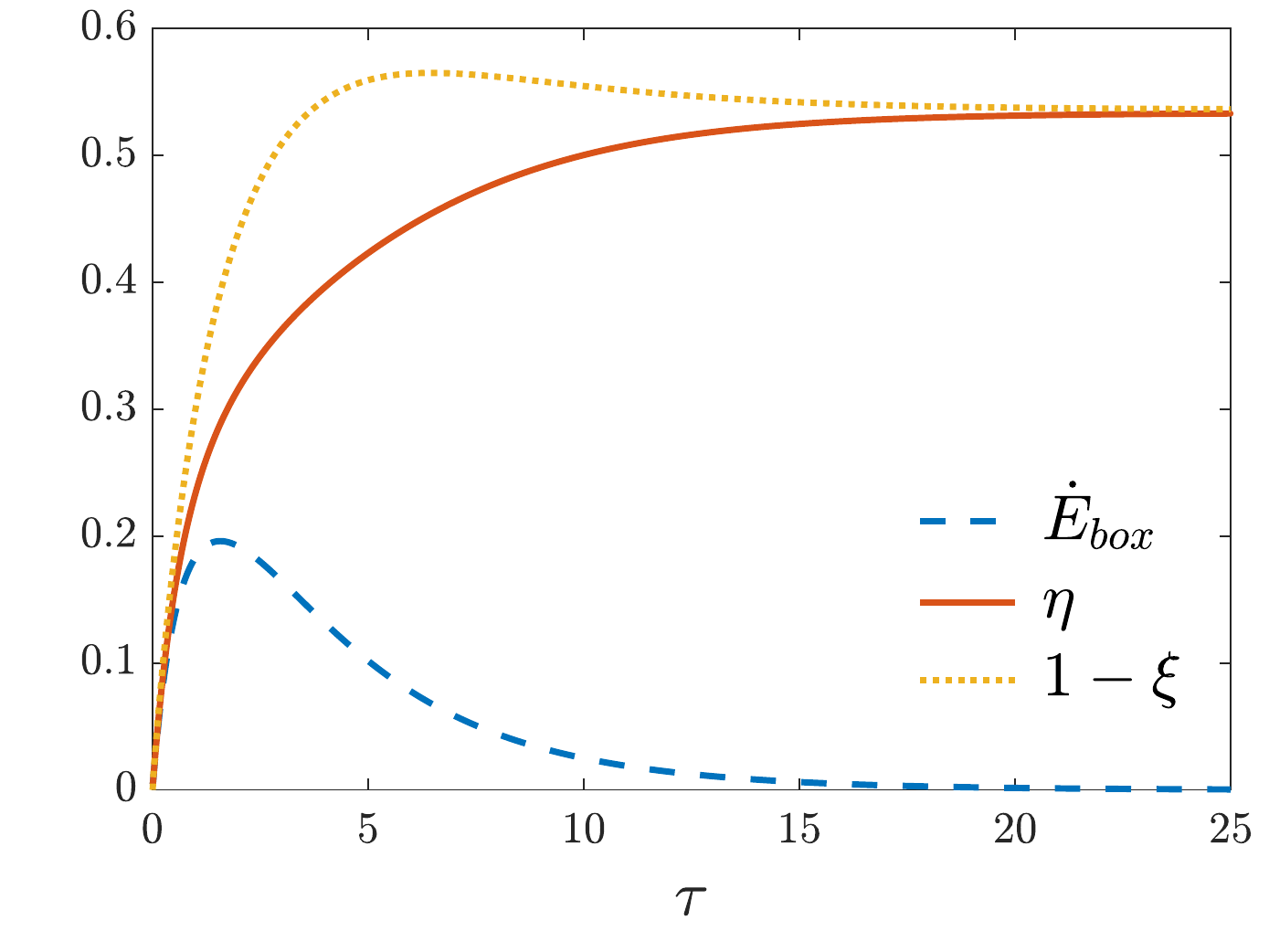}
		\caption{} \label{fig:emptyingfillingbox_efficiency}
	\end{subfigure}
	
	\caption{Emptying-filling box for $A^* C^{-3/2} H^{-2} = 0.02$: (a) Dimensionless density profile in an emptying-filling box at $\tau = 0.5, 1, 2, 5, 20$, (b) mixing efficiency of an emptying-filling box $\eta$ (solid, orange), normalised rate of change of potential energy inside the box $\dot{E}_{box}$ (dashed, blue) and the height of the mixed layer $1 - \xi$ (dotted, yellow) against non-dimensional time.} 
\end{figure}

The density profile for $\frac{A^*}{C^{3/2} H^2} = 0.02$ is plotted in Fig.\ \ref{fig:emptyingfillingbox_density}. A stratified layer grows until the stack pressure difference across the box is sufficient to drive a strong enough ventilation flow through the box to match the flow into the mixed layer by the plume. There is some overshoot of the mixed layer height. By $\tau = 20$ the box has essentially reached a steady state two-layer stratification.

We calculate the potential energy of the evolving stratification in the box $E_{box}$. The time rate of change $\dot{E}_{box}$ normalised by $\dot{E}_a$ is plotted in Fig.\ \ref{fig:emptyingfillingbox_efficiency} (dashed, blue). At early times, $\dot{E}_{box}$ is positive as the potential energy of the box grows with time. At $\tau \approx 2$ the rate of increase of potential energy reaches a peak. At late times the box is in steady state, therefore the potential energy of fluid in the box does not change with time. Our energetics analysis reveals that rate of change of potential energy of fluid in the box dominates the total potential energy change at early times and peaks before the height of the mixing region reaches its maximum.

The instantaneous mixing efficiency of the transient case can be calculated from the rate of change of potential energy in the box $\dot{E}_{box}$ and the rate of potential energy loss associated with the ventilated fluid $\dot{E}_{v}$. The total rate of change of background potential energy for the system is $\dot{E}_b = \dot{E}_{box} + \dot{E}_{v}$. The rate of potential energy loss due to ventilation is equal to the rate of potential energy increase if the ventilation flow was momentarily switched off in the model.

The instantaneous mixing efficiency is plotted in Fig.\ \ref{fig:emptyingfillingbox_efficiency} (solid, orange). The mixing efficiency increases monotonically up to a maximum value of $1 - \xi$. At early times the mixing efficiency is dominated by the growing stratification in the box, with little loss of background potential energy due to the through flow. At $\tau > 2$ the rate of change of background potential energy begins to be dominated by the ventilation flow rate. The mixing efficiency approaches $1 - \xi$ as the box approaches a steady state.

\section{Discussion}

The filling box has a maximum mixing efficiency of $\eta = \frac{1}{2}$, while for the emptying-filling box the maximum mixing efficiency $\eta = 1$. This difference in maximum mixing efficiency can be understood if we consider which part of the density profile is being altered by mixing.

For the filling box, the change in the density profile is equally distributed over the entire height. This change in density has a centre of mass at the mid-height of the box. In other words, buoyancy in the plume travels half-way down the box on average. This results in a mixing efficiency of $\eta = \frac{1}{2}$. Note that this is a result of the equal distribution of the density change in the asymptotic steady state and is independent of the exact shape of the density profile.

For the emptying-filling box, the steady state has a constant density in the mixed layer. The maximum mixing efficiency corresponds to the case when the mixed layer fills the box. The ventilation flow removes mixed fluid from the bottom of the box and adds unmixed fluid to the top of the box. To keep the box in steady state, buoyancy from the plume mixes with incoming ambient fluid to replenish the mixed layer. When the mixed layer fills the box, buoyancy from the plume remains near the top of the box. This results in a maximum mixing efficiency of $\eta = 1$.

The entrainment coefficient describes a rate and affects the timescale for mixing to occur. The mixing efficiency describes the energetic consequences of mixing. We can see this most clearly in the filling-box case, where the asymptotic steady state has a mixing efficiency of $\frac{1}{2}$, irrespective of the value of the entrainment coefficient, but the entrainment coefficient determines the time taken to reach the asymptotic steady state. 

For the case of the emptying-filling box, the mixing efficiency can be set by varying $A^*$, which is equivalent to varying the opening areas. We can interpret this as follows: by varying the flow rate through the box, we are changing the timescale over which mixing can occur. When $A^*$ is increased, fluid flows through the box at a faster rate and mixing has less time to occur before fluid exits the box. There is less mixing, therefore the mixing efficiency is smaller. On the other hand, this also illustrates another property of the mixing efficiency: it can be affected by an external parameter that is unrelated to the exact mechanism by which mixing occurs. Therefore, the value of the mixing efficiency does not necessarily reflect the fundamental mechanism behind turbulent entrainment.

\section{Acknowledgements}

The authors would like to thank Henry Burridge, Paul Linden, and Jeff Koseff for interesting and useful conversations.

\bibliographystyle{unsrt}
\bibliography{filling_box}

\end{document}